\documentclass[12pt]{article}

\usepackage{graphicx}

\newcommand{\beq}{\begin{equation}}
\newcommand{\eeq}{\end{equation}}
\newcommand{\beqa}{\begin{eqnarray}}
\newcommand{\eeqa}{\end{eqnarray}}
\newcommand{\beqar}{\begin{eqnarray*}}
\newcommand{\eeqar}{\end{eqnarray*}}

\begin{document}
\thispagestyle{empty}

\hfill{\sc UG-FT-213/07}

\vspace*{-2mm}
\hfill{\sc CAFPE-83/07}

\vspace{32pt}
\begin{center}
\centerline{\textbf{\Large
Little Higgs models with a light T quark}}

\vspace{40pt}

R. Barcel\'o, M. Masip and M.~Moreno-Torres$^*$
\vspace{12pt}

\textit{
CAFPE and Departamento de F{\'\i}sica Te\'orica y del
Cosmos}\\ \textit{Universidad de Granada, E-18071, Granada, Spain}\\
\vspace{16pt}
\texttt{barcelo@correo.ugr.es, masip@ugr.es, miguelmtt@ugr.es}
\end{center}

\vspace{40pt}

\date{\today}

\begin{abstract}

We study Little Higgs models based on a $SU(3)_1\times SU(3)_2$ 
global symmetry and with two scales (the two vacuum
expectation values $f_{1,2}$) substantially different.
We show that all the extra vector boson fields present
in these models
may be much heavier than the vectorlike $T$ quark 
necessary to cancel top-quark quadratic corrections. 
In this case the models become an extension of the 
standard model with a light ($\approx 500$ GeV) $T$ quark 
and a scalar Higgs field with a large singlet component. 
We obtain that the Yukawa and the gauge couplings 
of the Higgs are smaller than in the standard model,
a fact that may reduce significantly the Higgs production 
rate through glu-glu and $WW$ fusion. The $T$-quark 
decay into Higgs boson becomes then a dominant Higgs 
production channel in hadron colliders.

\end{abstract}

\vspace{20pt}

$^*$Present Address: {\it Depto.~de F\'\i sica Moderna, Univ.~de Granada,
18071 Granada, Spain.}

\newpage

\section{Introduction}

The {\it naturalness} of the electroweak (EW) scale has been
the main motivation to presume new physics below 
$\Lambda\approx 1$ TeV. We have hints 
of a large grand unification scale, and we know that the Planck scale
is there, so we {\it need} a mechanism that cancels
the quantum corrections introduced by these large scales. 
The majoritary point of view has been that supersymmetry,
technicolor or, more recently, extra dimensions could do
the job and rise the cutoff of the standard model (SM) up to the 
fundamental scale.

This point of view, however, has become increasingly uneasy 
when facing the experimental evidence. Flavor physics, electric 
dipole moments and other precision electroweak observables 
suggest that, if present, the sfermion masses, the technicolor
gauge bosons, or the Kaluza-Klein excitations of the standard
gauge fields are above 5 TeV \cite{Erler:2004nh}. 
To be effective below the TeV and consistent with the data
these models require a {\it per cent} 
fine tuning, whereas their presence at 5 TeV implies 
that nature deals with 
the Higgs mass parameter $m_h^2$ first using a mechanim to 
cancel 30 digits and then 
playing {\it hide and seek} with the last two digits.
It may be more consistent either to presume that 
there should be {\it another} reason explaining this 
{\it little hierarchy} between the EW and the scale of 
new physics or that there is no 
dynamical mechanism that cancels {\it any} fine tuning 
in $m_h^2$ \cite{Agrawal:1997gf,Arkani-Hamed:2004fb}. 
This second possibility has been
seriously considered after recent astrophysical and
cosmological data suggested a non-zero vacuum energy density 
(the preferred value does not seem to be explained by 
any dynamics at that scale), and it will be clearly 
favored if no physics beyond the SM is
observed at the LHC.

Little Higgs (LH) ideas
\cite{Georgi:1974yw,Arkani-Hamed:2001nc,Arkani-Hamed:2002qy}
provide a very interesting framework with natural 
cancelations in the scalar sector. 
New symmetries 
protect the EW scale and define consistent models with a
cutoff as high as $\Lambda\approx 10$ TeV, scale where a more
fundamental mechanism (SUSY \cite{Csaki:2005fc}
or extra dimensions \cite{Contino:2003ve}) would 
manifest. Therefore, 
these models could describe all the physics
to be explored in the next generation of accelerators.
More precisely,
in LH models the scalar sector has a (tree-level) global
symmetry that is broken spontaneously at a scale
$f\approx 1$ TeV. The SM Higgs doublet is then a Goldstone
boson (GB) of the broken symmetry, and
remains massless and with a flat potential
at that scale. Yukawa and gauge interactions break
explicitly the global symmetry. However, the models
are built in such a way that the loop diagrams giving non-symmetric
contributions must contain at least two different
couplings. This {\it collective} breaking keeps
the Higgs sector free of one-loop quadratic top-quark and
gauge contributions (see 
\cite{Perelstein:2005ka,Schmaltz:2005ky} for a recent review).
Two types of models have been extensively considered
in the literature: the ones based on a simple group 
({\it littlest} $SU(5)$ model \cite{Arkani-Hamed:2001nc}) 
and the ones based on a product
group ({\it simplest} $SU(3)\times SU(3)$ model
\cite{Schmaltz:2004de}) of global symmetries.

Of course, an important point is then if the
new physics that these models introduce is consistent 
with the data. In \cite{Casas:2005ev} it is shown that 
in general this is not the case, and the degree of 
fine tuning that they require is not below the one,
for example, in the MSSM. LH models include
an extra $T$ quark (that cancels quadratic top quark
corrections) and massive gauge boson fields 
(that cancel quadratic gauge contributions and absorb
extra GBs that otherwise would be massless).
It is this later type of fields, the extra gauge bosons, 
the one giving large corrections to EW 
observables through mixing with the standard gauge bosons 
and through direct couplings with the light fermions.
In models based on a global $SU(5)$ symmetry 
this problem can be solved imposing a $Z_2$ symmetry 
known as $T$ parity \cite{Cheng:2003ju,Cheng:2004yc}. 
This symmetry is 
analogous to the $R$ parity of SUSY models: under $T$ 
all the extra fields except for the $T$ quark are odd, whereas 
the standard fields are even. As a consequence, the
mixing of standard and extra gauge bosons as well as the 
tree level exchange of extra bosons by standard 
fermions are forbidden. This keeps the corrections
under control and allows $T$ quarks as low as 500 GeV,
as required for an effective cancelation of quadratic 
corrections in the Higgs sector. Unfortunately, in 
the {\it simplest} $SU(3)\times SU(3)$ model it is not 
possible to implement a $T$ parity.

In this paper we explore the simplest LH model 
\cite{Schmaltz:2004de,Kaplan:2003uc,Cheung:2006nk}
and find that it can also accommodate a relatively 
light $T$ quark together with suppressed extra gauge boson
contributions. This is achieved when the two VEVs $f_1$
and $f_2$ in these models are substantially different.
We analize this case and show that it has important 
phenomenological implications in Higgs physics.
As we will argue, analogous results would be obtained
in any LH model with a relatively low scale of global symmetry 
breaking.

\section{The model}
Let us start describing in some detail the 
model. The scalar sector 
contains two triplets, $\phi_1$ and $\phi_2$, of a global
$SU(3)_1\times SU(3)_2$ symmetry:
\beq
\phi_1\rightarrow e^{i\theta^a_1 T^a} \phi_1\;,\;\;\;\;
\phi_2\rightarrow e^{i\theta^a_2 T^a} \phi_2\;,
\eeq
where $T^a$ are the generators of $SU(3)$. 
It is assumed that the scalar triplets get vacuum expectation
values (VEVs) $f_{1,2}$ and break the global symmetry
to $SU(2)_1\times SU(2)_2$. The spectrum of scalar fields
at this scale consists then of 10 massless modes (the GBs of the 
broken global symmetry) plus two massive fields (with masses
of order $f_1$ and $f_2$). If one combination of the two 
global $SU(3)$ is made local, some of the 
GBs will be {\it eaten} by massive gauge bosons and the rest
will define the SM Higgs sector. 

In particular, if the two VEVs are 
\beq
\langle \phi_{1}\rangle = 
\left(\begin{array}{c} 0 \\ 0 \\ f_1  \end{array}\right)\; , \;\;\;\;
\langle \phi_{2}\rangle = 
\left(\begin{array}{c} 0 \\ 0 \\ f_2  \end{array}\right)\;,
\eeq
and the diagonal combination of $SU(3)_1\times SU(3)_2$ is made local,
\beq
\phi_{1}\rightarrow e^{i\theta^a(x) T^a} \phi_{1}\;,\;\;\;\;
\phi_{2}\rightarrow e^{i\theta^a(x) T^a} \phi_{2}\;,
\eeq
then the VEVs break $SU(3)\times U(1)_\chi$ to the standard
$SU(2)_L\times U(1)_Y$, a process that takes 5 GBs.
The other 5 GBs (the complex doublet $(h^0\; h^-)$ and a CP-odd 
singlet $\eta$) can be parametrized 
non-linearly \cite{Coleman:1969sm}:
\beq
\phi_{1}= e^{+ i\; {f_2\over f_1} \Theta }
\left(\begin{array}{c} 0 \\  0 \\ f_1
\end{array}\right)\;,\;\;\;\;
\phi_{2}= e^{- i\;{f_1\over f_2} \Theta}
\left(\begin{array}{c} 0 \\  0 \\ f_2
\end{array}\right)\;,
\label{paramet}
\eeq
where 
\beq
\Theta= {1\over f}\;
\left(\begin{array}{ccc} \eta/\sqrt{2} & 0 &h^0 \\
0& \eta/\sqrt{2} & h^- \\
h^{0\dagger} & h^+ & \eta /\sqrt{2} \end{array}\right) 
\label{theta}
\eeq
and $f=\sqrt{f_1^2+f_2^2}$.

If the global symmetry were exact, the Higgs boson would
be massless. However, the symmetry is just {\it approximate}
(it is broken by top-quark and gauge-boson loops), 
so we expect that the non-symmetric operators
will appear just suppressed by a loop factor.
This may (should) give an acceptable VEV and a mass 
to the Higgs \cite{Perelstein:2003wd}. 
Let us then assume that the real component of
$h^0$ gets a VEV, 
\beq
\langle h^0 \rangle = u/\sqrt{2}\;,
\eeq 
and gives
mass to the standard fermions and gauge bosons. This Higgs
VEV implies the triplet VEVs
\beq
\langle \phi_{1}\rangle = 
\left(\begin{array}{c} i f_1 s_1 \\ 0 \\ f_1 c_1  \end{array}\right)
\; , \;\;\;\;
\langle \phi_{2}\rangle = 
\left(\begin{array}{c} -i f_2 s_2 \\ 0 \\ f_2 c_2 \end{array}\right)\;,
\eeq
where
\beq
s_1\equiv \sin {u f_2\over \sqrt{2} f f_1}
\; , \;\;\;\;
s_2\equiv \sin {u f_1\over \sqrt{2} f f_2}
\;.
\eeq
To obtain the observed $W$ and $Z$ masses one needs
\beq
\sqrt{f_1^2 s_1^2 + f_2^2 s_2^2} = {v\over \sqrt{2}} = 
174\;{\rm GeV}
\;.
\eeq
Notice that using this non-linear realization of
$\phi_i$ the Higgs VEV $u$ is not 246 GeV, although it goes
to this value in the limit of large $f_1$ and $f_2$ 
(small $s_1$ and $s_2$).

In the unitary gauge all the GBs except for the singlet
$\eta$ and the 
standard neutral Higgs $h$ are eaten by massive gauge bosons.
In particular, it is easy to deduce the relation between 
$\phi_{1,2}$ and these fields:
\beqa
\phi_1&=& \exp\left( i{f_2 \eta\over f_1 f \sqrt{2}}\right)
\left(\begin{array}{c} i f_1 \displaystyle
(s_1\cos {h f_2\over \sqrt{2} f f_1}+c_1\sin {h f_2\over \sqrt{2} f f_1})\\ 
0 \\ 
f_1 \displaystyle
(c_1\cos {h f_2\over \sqrt{2} f f_1}-s_1\sin {h f_2\over \sqrt{2} f f_1})
\end{array}\right)\;,\nonumber \\
& & \nonumber \\
\phi_2&=& \exp\left( -i{f_1 \eta\over f_2 f \sqrt{2}}\right)
\left(\begin{array}{c} -i f_2 \displaystyle
(s_2\cos {h f_1\over \sqrt{2} f f_2}+c_2\sin {h f_1\over \sqrt{2} f f_2})\\ 
0 \\ 
f_2 \displaystyle
(c_2\cos {h f_1\over \sqrt{2} f f_2}-s_2\sin {h f_1\over \sqrt{2} f f_2})
\end{array}\right)\;.
\label{GBs}
\eeqa

\section{A light T quark}
As explained in the introduction, we are interested in models where
the extra gauge bosons are heavy 
while the vectorlike $T$ quark that cancels quadratic corrections
is lighter. The first requirement fixes the scale
$f=\sqrt{f_1^2+f_2^2}$, as the gauge boson masses are $\approx gf$. 
The top-quark Yukawa sector includes a triplet
$\Psi_Q^T=(t\; b\; T)$ and two singlets ($t^c_1$,
$t^c_2$), and it is described by the Lagrangian
\beq
-{\cal L}_t = \lambda_1\; \phi_1^\dagger \Psi_Q t_1^c + \lambda_2
\;\phi_2^\dagger \Psi_Q t_2^c + {\rm h.c.}\;,
\label{eq5}
\eeq
where all the fermion fields are two-component spinors.
In the limit of exact symmetry ({\it i.e.}, $s_1=0=s_2$) the
top quark is massless and the extra $T$ quark has a mass
$m_T$:
\beq
-{\cal L}_t \supset \lambda_1\; f_1 T t_1^c + \lambda_2\;
f_2 T t_2^c + {\rm h.c.} = m_T\; T T^c
 + {\rm h.c.}\;,
\label{eq6}
\eeq
where $m_T=\sqrt{\lambda_1^2 f_1^2 + \lambda_2^2 f_2^2}$ and  
$\;T^c= s_\alpha\; t_1^c+c_\alpha\; t_2^c$, with
$s_\alpha=\lambda_1 f_1/\sqrt{\lambda_1^2 f_1^2+\lambda_2^2 f_2^2}$. 

Once the Higgs VEV $u$ is included, it is easy to see that 
in order to have $m_T$ significantly smaller than $f$
we need $f_1\ll f_2$ and $\lambda_2\ll \lambda_1$.
If we define $f_1\equiv \epsilon f_2$, 
$\lambda_1 \equiv \lambda$ and $\lambda_2\equiv \epsilon' \lambda$
this means that $\epsilon$ and $\epsilon'$ are small. 
At first order in these two parameters we have $f_2\approx f$, 
$f_1\approx \epsilon f$ and 
\beqa
s_1&=&\sin {u f_2\over \sqrt{2} f f_1}\approx 
\sin{u\over \sqrt{2} \epsilon f}\;,\nonumber \\
s_2&=&\sin {u f_1\over \sqrt{2} f f_2}\approx 
{\epsilon u\over \sqrt{2} f}\;.
\eeqa
This implies (we redefine the top-quark field 
$-it\rightarrow t$)
\beq
-{\cal L}_t \supset \lambda \epsilon f \; (s_1 c_\alpha\; t t^c 
+ s_1 s_\alpha\; t T^c +
{c_1\over s_\alpha}\; T T^c)+ {\rm h.c.}
\;, 
\eeq
where $s_\alpha \equiv c_1\epsilon /\sqrt{c_1^2\epsilon^2 +
\epsilon'^2}$ and
\beq
t^c=c_\alpha\; t_1^c - s_\alpha\; t_2^c \;;\;\;\;\;
T^c=s_\alpha\; t_1^c + c_\alpha\; t_2^c \;.
\eeq
Taking $\epsilon f s_1\approx v/\sqrt{2}$ we
have
\beq
-{\cal L}_t \supset m_t\; t t^c
+ m_t t_\alpha\; t T^c +
{m_t\over c_\alpha s_\alpha t_1} T T^c+ {\rm h.c.}\;,
\eeq
with $m_t\approx \lambda v c_\alpha/\sqrt{2}$, 
$t_\alpha=s_\alpha/c_\alpha$ and 
$t_1=\tan (u/ \sqrt{2} \epsilon f)$.
To obtain the mass eigenstates (we denote them 
through the paper with a prime) we still need to
perform a rotation in the space of the left-handed 
fields $t$ and $T$:
\beq
t'=c_\theta\; t-s_\theta\; T \;;\;\;\;\;
T'=s_\theta\; t+c_\theta\; T \;,
\label{eigen}
\eeq
which imply a heavy mass and a mixing 
\beq
m_T\approx {m_t\over c_\alpha s_\alpha t_1}\;,\;\;\;\;
s_\theta\equiv V_{Tb}\approx {m_t t_\alpha\over m_T}\;.
\eeq

\section{Yukawa and gauge interactions}
It is now easy to find the approximate 
Higgs couplings with the top and 
the $T$ quark. At the lowest order in $\epsilon$
and $\epsilon'$ we obtain
\beq
-{\cal L}_t \supset
{m_t\over v} \left( c_1\; h t t^c
+ c_1 t_\alpha \; h t T^c - s_1 \; h T t^c -
s_1 t_\alpha \; h T T^c \right)+ {\rm h.c.}
\eeq
for the Yukawas and 
\beq
-{\cal L}_t \supset - {1\over 2 m_T} {m_t^2\over v^2} \left(
{s_1 c_1\over s_\alpha c_\alpha} \; h^2 t t^c
+ {s_1 c_1\over c_\alpha^2} \; h^2 t T^c 
+ {c_1^2\over s_\alpha c_\alpha} \; h^2 T t^c +
{c_1^2\over c_\alpha^2} \; h^2 T T^c \right) + {\rm h.c.}
\eeq
for the terms of dimension 5. 

This lagragian exhibits two features. The first one is common
to all LH models, namely, the quadratic corrections from
the diagrams in Fig.~1 cancel (the correction that these diagrams
introduce is logarithmic and proportional to $m_T^2$).
\begin{figure}
\begin{center}
\includegraphics[width=130mm]{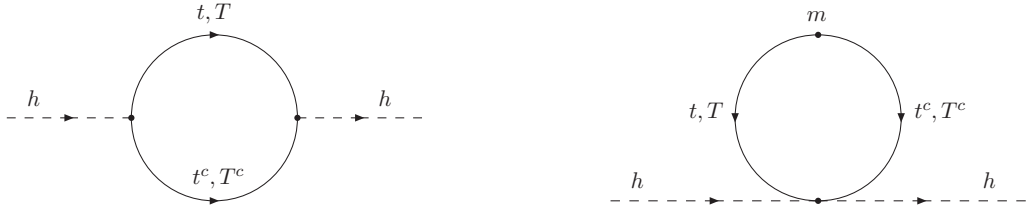}
\caption{1-loop corrections to $m_h^2$.
\label{f1}}
\end{center}
\end{figure}
The second one is that the top-quark Yukawa coupling $y_t$
is {\it not} $\sqrt{2} m_t/v$, like in the SM. Here the coupling 
appears suppressed by a factor of $c_1$. This is a generic feature 
in all LH models, and
it can be understood as the contribution  
of higher dimensional operators in
the non-linear expansion 
or as a mixing of order $v/(\sqrt{2}f_1)$ of the Higgs 
doublet with the $SU(2)_L$ singlets breaking the global symmetries. 
In the model under study the scale $f_1$ where 
$SU(3)_1$ is broken is relatively low, so the effect becomes
important. Notice that only the doublet component of the Higgs
couples to the fermions and gives them a mass 
$y^{SM}_f v/\sqrt{2}$. Then, if the doublet is just a 
component $c_1$ along the physical Higgs, the Yukawa 
couplings will be $y_f= y^{SM}_f c_1$.
Actually, in the model under study we still have to perform
the rotation in Eq.~(\ref{eigen}) to obtain the quark mass 
eigenstates. The flavor-diagonal Yukawa couplings are then
\beq
-{\cal L}_t \supset
{m_t\over v} \left( \left(c_1 c_\theta
+s_1 s_\theta \right) \; h t' {t^c}' -
\left( s_1 t_\alpha c_\theta - c_1 t_\alpha s_\theta\right)
\; h T' {T^c}' \right)+ {\rm h.c.}\;,
\eeq
which imply
\beq
{y_t\over y^{SM}_t}\approx c_1 c_\theta +s_1 s_\theta
\approx c_1 + s_1 V_{Tb}\; . 
\label{yuk}
\eeq
In the gauge sector we obtain the same type of suppression
effect. The gauge couplings of the Higgs $h$ 
with both the $W$ and the $Z$ vector bosons are reduced with 
respect to the SM values by a factor of 
\beq
{g\over g^{SM}}= {\sqrt{2} f_1 f_2 (s_1 c_1+s_2 c_2)\over v f}
\approx c_1\; , 
\label{gauge}
\eeq
where $g$ stands for the $SU(2)_L$ and $U(1)_Y$ couplings.
In this model the singlet component $s_1$ along the Higgs,
the mass $m_T$ of the vectorlike quark, and the mixing 
$V_{Tb}$ between $T$ and $t$ depend only on two parameters 
($\epsilon$ and $\epsilon'$), which yields the approximate 
relation
\beq
{s_1\over c_1} \approx V_{Tb}+ {m_t^2\over m_T^2 V_{Tb}}\;. 
\eeq
In Fig.~2 we plot the exact (numerical) correlation
among these three quantities.
\begin{figure}
\begin{center}
\includegraphics[width=110mm]{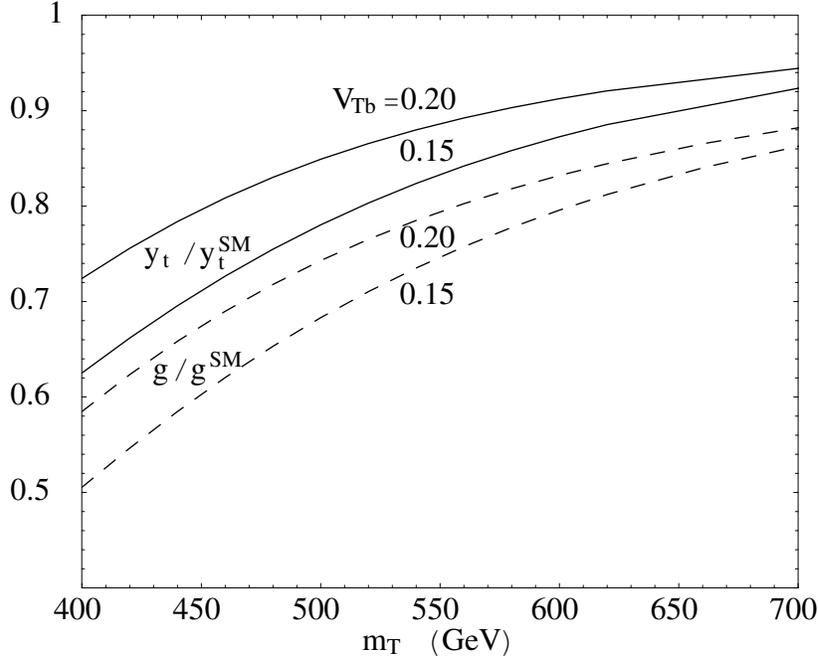}
\caption{Suppresion of the top quark (solid) and the gauge 
(dashes) couplings versus the SM values for 
$V_{Tb}=0.20,0.15$ and different values of $m_T$.
\label{f2}}
\end{center}
\end{figure}

The approximate Yukawa couplings with the (pseudo) GB $\eta$ 
can also be obtained expanding $\phi_{1,2}$ in Eq.~(\ref{GBs}):
\beq
-{\cal L}_t \supset
i{m_t\over v} \left( s_1\; \eta t t^c
+ s_1 t_\alpha \; \eta t T^c + c_1 \; \eta T t^c +
c_1 t_\alpha\; \eta T T^c \right)+ {\rm h.c.}
\eeq

It is also easy to deduce the couplings of the
top and the $T'$ quarks with the gauge bosons. 
After the rotation in Eq.~(\ref{eigen})
and neglecting
the mixings with the lighter quarks,
the couplings with the $W$ boson are
\beqa
{\cal L}_W &=& -{g\over \sqrt{2}}\;
\overline t \sigma^\mu b \;
W^+_\mu + {\rm h.c.}\nonumber \\
&=& -{g\over \sqrt{2}}
\left( \sqrt{1-V^2_{Tb}} \;\overline t'\sigma^\mu b 
+ V_{Tb}\; \overline T'\sigma^\mu b \right)
W^+_\mu + {\rm h.c.} 
\eeqa
In the $Z$-boson sector we
find additional flavour-changing interactions:
\beqa
{\cal L}_Z &\supset & -{g\over 2 c_W}
\left( {\begin{array}{cc}
\overline t & \overline T \\
\end{array}} \right)
\sigma^\mu
\left( {\begin{array}{cc}
1 & 0 \\
0 & 0 \\
\end{array}} \right)
\left( {\begin{array}{c}
t \\
T \\
\end{array}} \right)
Z_\mu \nonumber \\
&\approx &
-{g\over 2 c_W}
\left( {\begin{array}{cc}
\overline t' & \overline T' \\
\end{array}} \right)
\sigma^\mu
\left( {\begin{array}{cc}
1-V^2_{Tb} & V_{Tb} \\
V_{Tb} & V^2_{Tb} \\
\end{array}} \right)
\left( {\begin{array}{c}
t' \\
T' \\
\end{array}} \right)
Z_\mu \;,
\eeqa
{\it i.e.,} $X_{Tt}= V_{Tb}$ (see \cite{Aguilar-Saavedra:2002kr} 
for notation).

To be more definite, let us write a numerical example with
$f_1=300$ GeV, $f_2=4$ TeV, and $\lambda_2/\lambda_1=0.1$. Once
the Higgs gets a VEV with $s_1=0.5$ ($s_2=0.005$) and gives
mass to the EW gauge bosons, we obtain the standard $t'$ quark
plus a $T'$ state of mass $m_T=574$ GeV.  
The mass eigenstates are
\beqa
&&t'= 0.98\; t - 0.20\; T\;;\;\;\;\;
{t^c}'= 0.81\; t_1^c - 0.57\; t_2^c\; ,\nonumber \\ 
&&T'= 0.20\; t + 0.98\; T\;;\;\;\;\;
{T^c}'= 0.57\; t_1^c + 0.81\; t_2^c\; ,
\eeqa
which imply $V_{Tb}= 0.20$.
The dimension four and five couplings
of these fields with the Higgs read
\beq
-{\cal L}_t \supset
{m_t\over v} \left( 0.89\; h t' {t^c}'
+ 0.62 \; h t' {T^c}' - 0.38 \; h T' {t^c}' -
0.27 \; h T' {T^c}' \right)+ {\rm h.c.}
\eeq
and 
\beq
-{\cal L}_t \supset - {1\over 2 M_T} {m_t^2\over v^2} \left(
0.73\; h^2 t' {t^c}'
+ 0.51 \; h^2 t' {T^c}' 
 + 1.69 \; h^2 T' {t^c}' +
1.19 \; h^2 T' {T^c}' \right) + {\rm h.c.}
\eeq
This means that
$y_t/y_t^{SM}= 0.89$. 
The Yukawa couplings with $\eta$ are
\beq
-{\cal L}_t \supset
i{m_t\over v} \left( 0.38\; \eta t' {t^c}'
+ 0.27 \; \eta t' {T^c}' + 0.89 \; \eta T' {t^c}' +
0.62 \; \eta T' {T^c}' \right)+ {\rm h.c.}\;,
\eeq
whereas the couplings of these quarks with the $W$ boson are
\beq
{\cal L}_W = -{g\over \sqrt{2}}
\left( 0.98 \;\overline t' b 
+ 0.20 \;\overline T' b \right)
\gamma^\mu W^+_\mu + {\rm h.c.} 
\eeq
The interactions with the $Z$ boson
include the flavour-changing terms
\beq
{\cal L}_Z \supset  -{g\over 2 c_W}
\left( {\begin{array}{cc}
\overline t' & \overline T' \\
\end{array}} \right)
\sigma^\mu
\left( {\begin{array}{cc}
0.96 & 0.20 \\
0.20 & 0.04 \\
\end{array}} \right)
\left( {\begin{array}{c}
t' \\
T' \\
\end{array}} \right)
Z_\mu \;,\\
\eeq
It is remarkable 
that in models with just a vectorlike $T$ quark 
the mass eigenstates couple to the Higgs only 
through the term that also introduces the mixing with 
the top quark, {\it i.e.}, with a coupling 
$V_{Tb} m_T/v$ \cite{Aguilar-Saavedra:2002kr}. 
Here, however, the Higgs couples to $T$ and
$T^c$ even if the mixing $V_{Tb}$ is zero.  

Finally, 
several comments about the {\it stability} of the
scales are in order. First, notice that the natural 
cutoff of this model is at 
$\Lambda\approx 4\pi f_1$.
In the limit of $f_1=v/\sqrt{2}$ (i.e., $s_1=1$) this
is just the SM cutoff, whereas values of 
$f_1$ around 300 GeV rise the cutoff by a factor 
of 2 up to $\Lambda \approx 4$ TeV. 
Second, in order to decouple the extra gauge bosons
we are taking a large $f_2$ scale, around
the cutoff $\Lambda$. 
This defines a {\it minimal}
LH model with only a $T$ quark at 500 GeV that cancels 
top quark corrections, a light\footnote{Being a gauge 
singlet, $\eta$  avoids LEP bounds.}
singlet $\eta$, and a light Higgs $h$ whose
coupling with the Z boson is suppressed by a factor of
$c_1$. This suppression should relax LEP bounds on its
mass, which tend to be below 100 GeV. 

\section{Electroweak precision observables}
Let us start analyzing the implications on EW precision 
observables. 
The three basic ingredients of the LH models under study
are the presence of heavy vector bosons, of a relatively
light $T$ quark, and of a sizeable singlet component in the
Higgs field. 

{\it (i)} The massive gauge bosons would introduce mixing 
with the standard bosons and four fermion operators. This
could manifest as a shift in the $Z$ mass and corrections 
in atomic parity violation experiments and LEP II data. 
However, none of these effects is observable 
if $f_2 \ge 3$ TeV \cite{Schmaltz:2004de}).

{\it (ii)} The effects on EW precision observables due to 
the singlet component of the Higgs field are also 
negligible. 
Although the Yukawa coupling of the top 
with the neutral Higgs is here smaller than in the SM, it is
the coupling with the {\it would be} GBs (the scalars 
eaten by the $W$ and $Z$ bosons) what determines
the large top-quark radiative corrections, and these 
are not affected by the presence of singlets. 

{\it (iii)} The bounds on a vectorlike $T$ quark from
precision EW data  
have been extensively studied in the literature, 
we will comment here the results in 
\cite{Aguilar-Saavedra:2002kr} as they apply
to LH models in a straightforward way.

The mixing of the top quark with the $T$ singlet reduces
its coupling with the $Z$ boson. This, in turn, affects the top 
quark radiative corrections (triangle diagrams) to the $Zbb$ 
vertex, which is measured in the partial width $Z$ to  
$b\overline b$ [$R_b=\Gamma(Z\rightarrow b\overline b)/
\Gamma(Z\rightarrow {\rm hadrons})$]
and forward-backward asymmetries.
The heavier $T$ quark also gives this 
type of corrections to the $Zbb$ 
vertex, and for low values of $m_T$ 
both effects tend to 
cancell ({\it i.e.}, if $m_T=m_t$ the vertex $Zbb$ 
is the same as in the SM). For large values of $m_T$ (above
500 GeV) the upper bound on $V_{Tb}$ from precision $b$ 
physics is around 0.2 \cite{Aguilar-Saavedra:2002kr}.

The $T$ quark would also appear in vacuum polarization
diagrams, affecting the oblique parameters $S$, $T$, and 
$U$. For degenerate masses ($m_T=m_t$) the corrections to
$T$ and $U$ vanish for any value of the mixing $V_{Tb}$ 
and the correction to $S$ is small 
($\Delta S\approx -0.16 V_{Tb}$). 
For large values of 
$m_T$ the only oblique parameter with a 
sizeable correction is $T$ ($\Delta T\approx 2.7 V_{Tb}$ for 
$m_T=500$ GeV), but the limits on $V_{Tb}$ 
are in this case smaller than
the ones from $R_b$ \cite{Aguilar-Saavedra:2002kr}.

\section{Higgs physics}
The phenomenological impact of these models on Higgs physics
at hadron colliders 
may be important. The main effects can be summarized as follows.

{\it (i)} Suppression of the $gg\rightarrow h$ cross section.
This effect is due to the suppression of the top 
Yukawa coupling relative to  the SM value (see Fig.~2)
and also to the 
contribution of the extra $T$ quark. 
Although this second
factor is numerically less important, it is remarcable that
always interferes destructively in the amplitude: the relative
minus sign versus the top-quark contribution follows from the
cancelation of quadratic corrections to $m_h^2$.
Notice that the two diagrams for $gg\rightarrow h$ are obtained from
the diagrams in Fig.~1 just by adding two gluons to the fermion
loop and changing a Higgs leg by its VEV.  

It is easy to obtain approximate expressions for this suppression
factor in the limit of $m_H\ll m_t,m_T$ \cite{Rizzo:1979mf}:
\beqa
{\sigma(gg\rightarrow h)\over \sigma^{SM}(gg\rightarrow h)}
& \approx & \left( {y_t\over y_t^{SM}}+{y_T v\over m_T}\right)^2
\nonumber \\
& \approx & \left( c_1 c_\theta + s_1 s_\theta
-t_1 s^2_\alpha (s_1 c_\theta-c_1 s_\theta)\right)^2
\nonumber \\
& \approx & c_1^2
\eeqa
\begin{figure}
\begin{center}
\includegraphics[width=110mm]{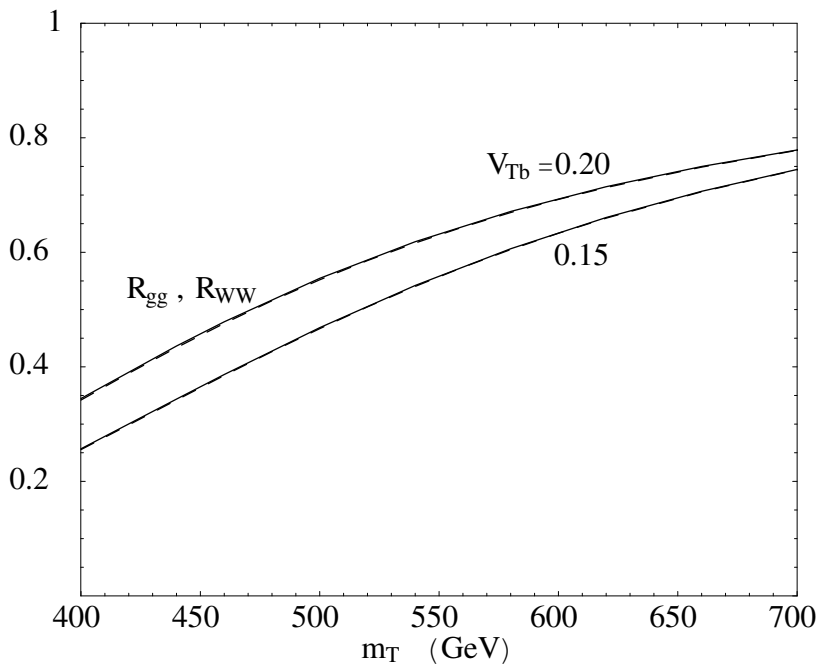}
\caption{Ratios $R_{gg}\equiv\sigma(gg\rightarrow h)/
\sigma^{SM}(gg\rightarrow h)$ (solid)
and $R_{WW}\equiv
\sigma(WW\rightarrow h)/\sigma^{SM}(WW\rightarrow h)$
(dashes) for 
$V_{Tb}=0.20,0.15$ and different values of $m_T$.
$R_{gg}$ and $R_{WW}$ coincide at the $1\%$ level.
\label{f3}}
\end{center}
\end{figure}
In Fig.~3 we plot the ratio 
$R_{gg}\equiv\sigma(gg\rightarrow h)/\sigma^{SM}(gg\rightarrow h)$
for different values of $V_{Tb}$ and $m_T$. For $m_H=150$ GeV
the approximation above is good at the $1\%$ level.
This effect, which could {\it hide} the Higgs at the LHC, 
has been recently discussed in general models with scalar
singlets \cite{O'Connell:2006wi,Bahat-Treidel:2006kx} and 
also in the framework 
of LH models with $T$ parity \cite{Chen:2006cs}.

{\it (ii)} Suppression in the production cross sections 
that involve gauge interactions: 
$WW\rightarrow h$, $q \overline q\rightarrow W h$, etc. 
[see Eq.~(\ref{gauge})].
We plot the ratio $R_{WW}\equiv
\sigma(WW\rightarrow h)/\sigma^{SM}(WW\rightarrow h)
\approx c_1^2$ also in Fig~3.
It is remarkable that for $m_H=150$ GeV the suppression
in these cross sections coincides with the one in
$\sigma(gg\rightarrow h)$ at the $1\%$ level.

{\it (iii)} New production channels through $T$-quark 
decay \cite{Han:2005ru}.
A $T$ quark of mass below 600 GeV will be copiously produced
at the LHC. In particular, the cross section to produce 
$T\overline T$ pairs in $pp$ collisions goes from 
$10^4$ fb for $m_T=400$ GeV to $10^3$ fb for $m_T=600$ GeV
\cite{Aguilar-Saavedra:2006gw}.
Once produced, a $T$ quark may decay into $Wb$, $Zt$, $ht$, and
$\eta t$ \cite{Kilian:2004pp}. 
We find an approximate relation among the
partial widths in the limit of $m_T$ much larger than the 
mass of the final particles:
\beqa
&&\Gamma(T\rightarrow Wb)\approx {\alpha\over 16 s_W^2}\;
V_{Tb}^2 \;{m_T^3\over M_W^2}\nonumber \\ 
&&\Gamma(T\rightarrow Zt)\approx 
{1\over 2}\;\Gamma(T\rightarrow Wb)\nonumber \\ 
&&\Gamma(T\rightarrow ht)\approx  {1\over 2}\;
(c_1^2+{s_1^2\over t_\alpha^2})\;
\Gamma(T\rightarrow Wb)\nonumber \\ 
&&\Gamma(T\rightarrow \eta t)\approx  {1\over 2}\;
(s_1^2+{c_1^2\over t_\alpha^2})\; \Gamma(T\rightarrow Wb)
\eeqa
Notice that the $T$ quark will decay through the 4 channels
with branching ratios that are independent of $V_{Tb}$.
$T\rightarrow W^+ b$ gives the best discovery potential 
for the $T$ quark, whereas the Higgs $h$ will be produced
with a branching ratio close to the $20\%$. The detailed 
signal and background study at the LHC 
in \cite{Aguilar-Saavedra:2006gw} 
shows that $T\overline T
\rightarrow W^+ b \overline t h\rightarrow W^+b W^- \overline b h$ 
and $T\overline T \rightarrow h t h \overline t
\rightarrow W^+ b W^- \overline b hh$
give a very high statistical significance for the Higgs
(around $10\sigma$ for 30 fb$^{-1}$). We expect similar
results in this model, although the presence of the scalar 
$\eta$ can open new decay channels for the Higgs.
In particular, if $m_H>2m_\eta$ the (global symmetry-breaking)
coupling $h\eta\eta$ could loosen LEP bounds on the
Higgs mass 
and open the interesting channel 
$h\rightarrow \eta\eta \rightarrow 4b$ \cite{Cheung:2006nk}.
In particular, if $m_H>2m_\eta$ the 
coupling $h\eta\eta$ opens the interesting channel 
$h\rightarrow \eta\eta \rightarrow 4b$ \cite{Cheung:2006nk}
that, together with the suppression in the $hZZ$ coupling, could
loosen considerably the 114 GeV LEP bound \cite{Barate:2003sz}
on the Higgs mass.

\section{Summary and discussion}
LH models are minimal extensions of the SM that rise
its {\it natural} cutoff. All these models 
contain a vectorlike $T$ quark that cancels one-loop 
quadratic corrections to $m_h^2$. An effective 
cancellation requires $m_T\approx 500$ GeV, which implies
a scale of global symmetry breaking of the same order. 
We have studied models based on a $SU(3)_1\times SU(3)_2$
global symmetry and have shown that all the other ingredients
of the models (namely, the extra vector and scalar fields)
can be decoupled. In LH models based on a simple
group this decoupling effect is achieved using a discrete 
symmetry known as $T$ parity, whereas here it is obtained 
fixing one of the VEVs ($f_2$) around 4 TeV and making
the other one ($f_1$) up to a factor of $4\pi$ smaller.

The Higgs $h$ has then suppressed gauge and Yukawa  
couplings. This effect can be understood as a mixing
of order $v/(\sqrt{2} f_1)$ of the doublet with the
singlet ($\sigma$) that breaks the global symmetries and
gives mass to the $T$ quark. 
This seems to be a generic feature in any LH models: the 
lighter 
is the extra $T$ quark that cancels top-quark corrections, 
the larger is the singlet component $s_1$ along the Higgs $h$.
In our model 
the scalar $\sigma$ that mixes with the doublet gets a mass
of order $f_1$. Notice that this 
scalar is necessary to unitarize the theory, since the
gauge coupling of the light Higgs $h$ is here suppressed 
by a factor of $c_1$ (if its mass is around the cutoff
$4\pi f_1$, in the limit $s_1=1$ the model becomes 
{\it Higgsless}).

The reduction in gauge and Yukawa
couplings respect to the SM values have consequences at
hadron colliders. In particular, the Higgs production 
rate through $gg$ and $WW$ fusion will be suppressed by
a factor of $c_1^2$. These models
are, actually, a realization of the ideas discussed in 
\cite{Patt:2006fw}, where the Higgs field is {\it spread} into 
several weaker modes. It is obvious that
this, together with the possible new decay 
mode $h\rightarrow \eta \eta$, could loosen substantially 
present bounds on the Higgs mass.

Although the standard channels to produce Higgs bosons 
are suppressed, the presence of a relatively light 
$T$ quark opens new possibilities. These fields will 
be copiously produced through tree-level interactions 
in hadron colliders, and they may decay into Higgs plus
a top quark. General analysis that can be found in the 
literature do not consider the effect of a
scalar singlet, as the mass of the quark is 
introduced {\it ad hoc} and not through scalar VEVs.
We have taken the singlet into account and have shown
that this decay mode has 
an approximate branching ratio of
the 20\% (versus the 25\% in models with no singlets). 

It is amusing that in these LH models the discovery of 
the Higgs at the LHC comes together
with the dicovery of a vectorlike quark and, thus, of a new
scale in particle physics (notice that $T$ is {\it not} 
at the EW scale). If that
were the case, the new {\it natural} cutoff of the model would 
be rised up to energies just above the reach of the LHC,
which would certainly provide for good 
arguments to build a {\it bigger} collider.

\section*{Acknowledgments}
We would like to thank Paco del \'Aguila and 
Juan Antonio Aguilar-Saavedra for useful discussions.
This work has been partially supported by
MEC (FPA2006-05294) and by Junta de Andaluc\'{\i}a
(FQM 101 and FQM 437). RB acknowledges a fellowship from
the Universidad de Granada.

\end{document}